# Associative control processor with a rigid structure

Magomedov I.A, Khazamov O.A

department of Computer Science, Dagestan State Technical University, Makhachkala city, 367014

**Abstract**

The approach of applying associative processor for decision making problem was proposed. It focuses on hardware implementations of fuzzy processing systems, associativity as effective management basis of fuzzy processor. The structural approach is being developed resulting in a quite simple and compact parallel associative memory unit (PAMU). The memory cost and speed comparison of processors with rigid and soft-variable structure is given. Also the example PAMU flashing is considered.

Key words: fuzzy control, fuzzy processor, decision making,

1. **Introduction**

An important place among various real-world problems is occupied by problems of developing logical-linguistic control models for complex technical systems[1]; i.e in the case study of applying logical-linguistic control model for vehicle crashworthiness modeling [2]. The solution can generally be reduced to generating descriptions of identifiable situations and constructing procedures for identifying these situations.

Resolving such problems, using a sequential computer, requires a considerable amount of time and a large RAM capacity, mainly due to the associative nature of identification procedures that require parallel performance of homogenous operations with a large set of data in real time.

## 2. Structural approach to implementation

The main idea of realizing fuzzy control algorithms consists in the set of possible solutions breaking the set of reference fuzzy situations into categories $\omega_k$, (k=1,...,K).

Each category may contain a significant number of reference situations $S_l^R$, (l=1,...,L) that match one solution.

To assign the current situation to a particular class one should determine the degree of membership $\mu(S^T, \omega_k)$ of the current fuzzy situation for each of these classes. In order to make a real-time decision, it is necessary to compare simultaneously the current fuzzy situation $S^C$ with all the reference fuzzy situations $S_l^R$ that already have pre-determined solutions.

Although there is an effective algorithm hierarchical graph neuron (HGN), which does not require definition of rules or setting of thresholds by the operator for the pattern recognition [3], following realization of algorithm for technical system fuzzy control models is suitable for real-time applications. It is reduced to classification tasks and their interpretation can be viewed from the view of structural approach to pattern recognition [4,5].

This approach consists in analyzing the input information structure in order to single out a set of attributes, in selecting decision rules and an efficient decision-making procedure that shall take into consideration correlations between the singled-out attributes, thus, categorizing the data and the input information in the aggregate.

It is believed that the set of attributes forms the terminal vocabulary, decision-making procedure forms the grammar (syntax), and the set of types, names, images, complexes of actions forms - the semantics or, in the terminology of [6], the semantic definition of a language model used to formally describe the behavior of complex system. A structural approach assumes:

1. Description of some fuzzy situation represented by a chain of attributes $S^T = (x_1, x_2, ..., x_i, ... x_n)$, prepared according to the rules of some grammar G.
2. Separate grammar $G_k$ is constructed for each class of situation $\omega_k$
3. If the chain $(x_1, x_2, ..., x_i, ... x_n)$, describing some situation is accepted by some grammar Gp, that is $S^T \in Q(G_P)$, then this situation belongs to a class $\omega_P$.

   The method «comparison with etalon» has three options: full coincidence, maximum resemblance and minimal difference.

   The option "full coincidence" is defined as follows :

   $S^T \sim \omega_P = \arg\{\Lambda^n_{i=1}(x_i \equiv S_{ik})\}$.

   That means recognizing situation ST defined by the chain $(x_1, x_2,..., x_i,...x_n)$ belongs to the class $\omega_P$ determined by etalon describing $S_P = (S_{1P} S_{2P}...S_{iP}...S_{nP})$ if all elements of the input descriptions $S_T$ coincide with the corresponding elements of the etalon description $S_P$ [5_7]

1. Let's define binary variables $S_i^j$, $\alpha_i^j$ as follows: ?

   $S_i^j =1$, if all incoming symbols from the 1-st to j-th inclusive belongs to the i-th description, $S_i^j =0$ otherwise.

   $\alpha_i^j =1$, if the j-th symbol of incoming sequence coincides with the j-th symbol of the i-th description, $\alpha_i^j =0$, otherwise.

   Under the symbol $\alpha_i^j$ in this case is meant any $\Psi(S)$ attribute.

2. Let's make connection between variables $S_i^j$ with following recurrent correlation: $S_i^j = S_{i-1}^j * \alpha_i^j, S_0 = 1$ ? (1)
3. Define $\delta$ and $\lambda$ functions of indicator.

The match of incoming word with i-th description is expressed by the fact that conjunction $S_j = S_1^j * S_2^j * \ldots * \alpha_{ni}^j$ equals to 1, (2) where $n_i$ is a number of characters in the i-th description.

Using (1) and (2) we can get $S^j = \alpha_1^j \alpha_2^j \ldots \alpha_{ni}^j$ \qquad (3)

Transform the expression (3) taking into account the values being formed at the given time points $\tau_1, \tau_2, \ldots, \tau_{ni}$.

$$S^j = \overline{\overline{\alpha_1^j \tau_1} \vee \overline{\alpha_2^j \tau_2} \vee \ldots \vee \overline{\alpha_{ni}^j \tau_{ni}}},$$ where $\alpha_I$ is an binary variable equal to 1 at time moment $\tau_j$.

Let's introduce signals $b(\tau_j) \equiv 1$, $c(\tau_0) \equiv 1$, $d(\tau_{ni} + \Delta\tau) \equiv 1$.

Then,

$$S^j = \overline{\overline{cb\alpha_1^j \tau_1} \vee \ldots \vee \overline{b\alpha_{ni}^j \tau_{ni} d}}, \qquad (4)$$

The expression (4) defines $\lambda$-function of indicator and the expression (1) defines $\delta$ function of indicator.

4. Proceeding from (4) obtain the scheme of indicator (fig. 1), that works as following:
   a) At the time moment $\tau_0$ the trigger is set to 1 by a signal c;
   b) At each time moment $\tau_j$ signal b and prohibiting its signal $\alpha_i^j$;
   c) At the time moment $(\tau_{ni} + \Delta\tau)$ indicator is requested by the signal d, being formed by signal "end of incoming characters"
   d) Outgoing signal $S^j$ polls corresponding cell of parallel associative memory device (PAMU).

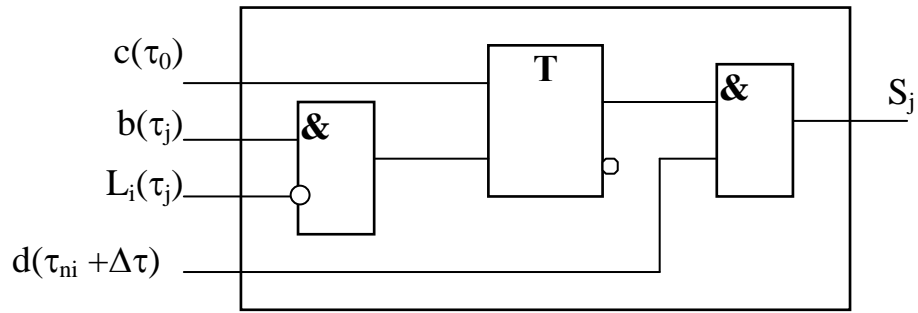

**Fig. 1 Functional schema of indicator**

## 3. Alternative proposals

The design of processor, shown on fig. 1a adds flexibility, since it's proposed with pipelined parallelism and the possibility to expand the capacity by simply adding other modules [8, 9].

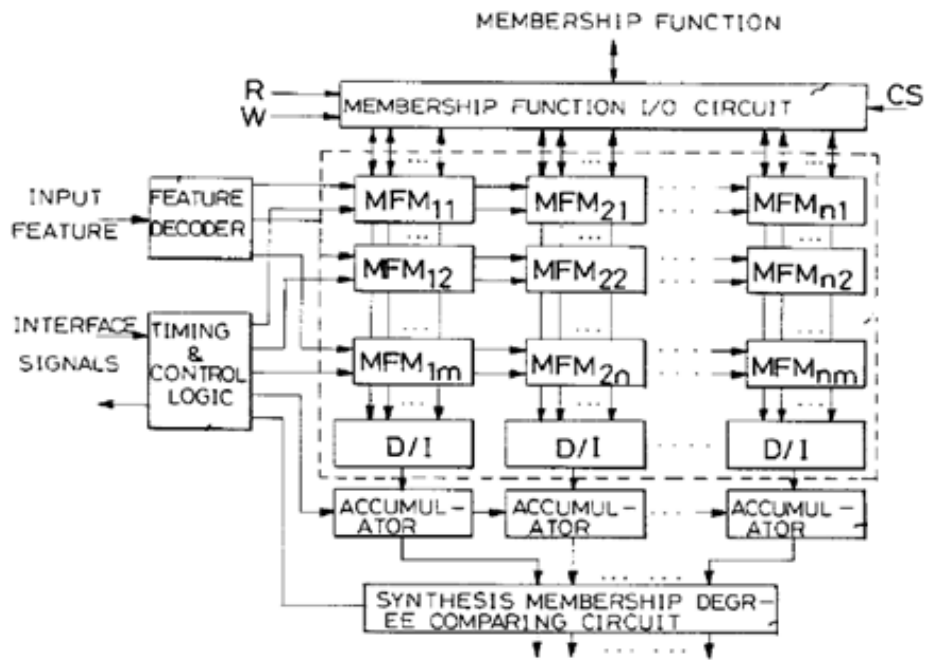

**Fig. 1a**

But the additional flexibility is gained in the expense of time. Consider the calculation of speed of the associative processor with flexible structure, given with its general block diagram shown on the fig. 2a.

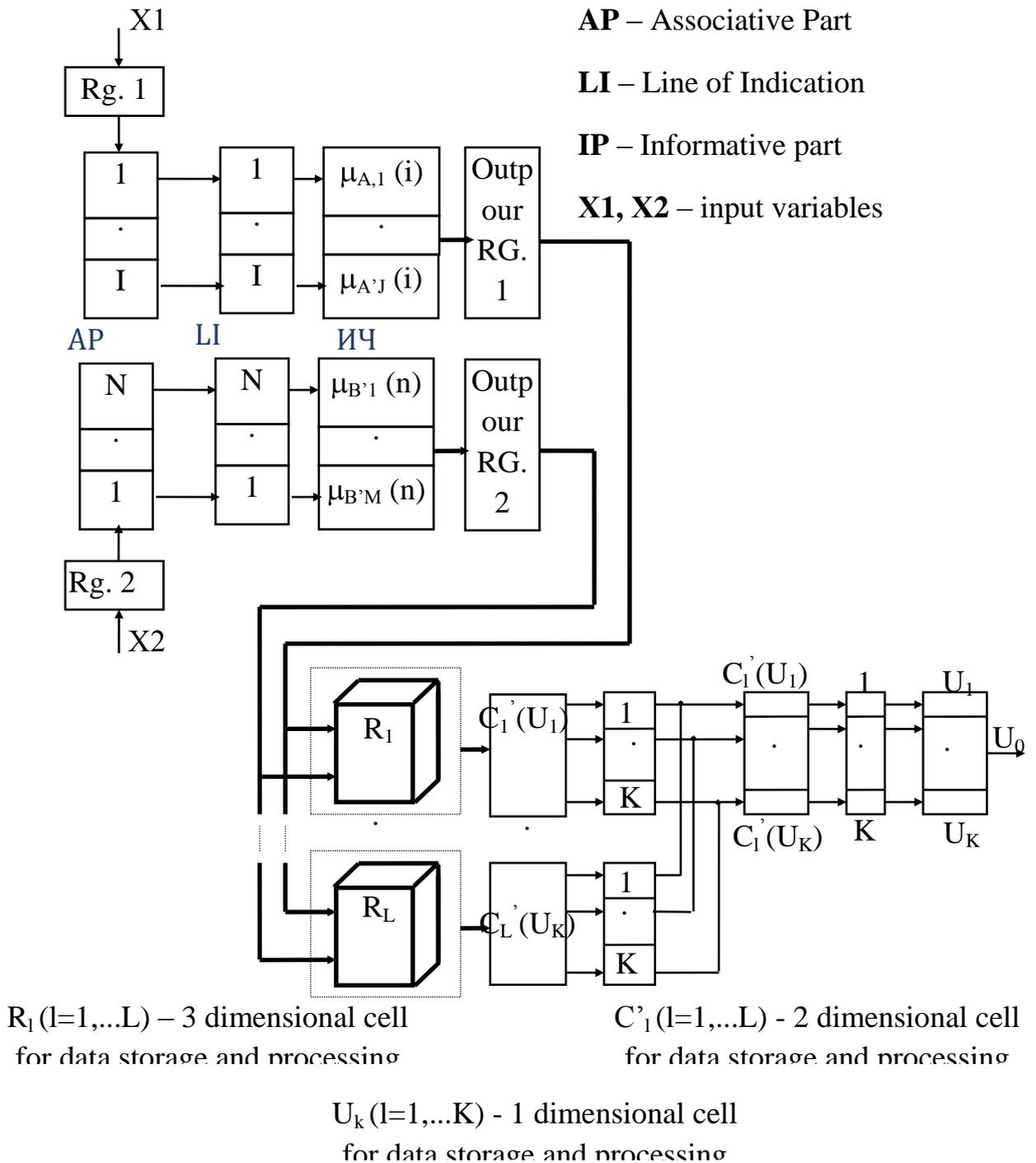

**Fig. 2a** Block diagram of the associative processor with flexible structure

Average processing time of the input data stream for the control task in the associative processor with flexible can be determined by the formula

$$T_c = T_{LC} + T_{fr} + T_{cs} = t_c + 4t_l + t_n^c + t^{min}_n + (n+1)t^{max}_n = \tau(6 + n\gamma + 2\gamma) \quad (5)$$

Where $T_{LC}$ is the time required for conversion of numeric data into fuzzy sets in the fuzzyfication unit ; $T_{fr}$ is the time for assertion of fuzzy rules; $T_{cs}$ is the time of the control signal forming; $t_c, t_l, t_n^c, t^{min}_n, t^{max}_n$ – the time of reading of information from the coordinate block, seeking information in the fuzzyfication unit, data fetching from the fuzzyfication unit, rule base block and control signal block accordingly; n – number of input variables, γ - width of processing words, τ - duration of control signal.

It can be concluded from (5), that the time depends upon the duration of tact signals, capacity of processing words and the number of input variables.

The memory cost is evaluated as follows:

$$V_\Pi = \sum_{n=1}^{N}(V_{LC} + V_{fr}) + V_{CS} = \sum_{n=1}^{N}(V_{AP} + V_{IP} + V_k) + V_{cs} = \sum_{n=1}^{N}(\gamma I_n + I_n + I_n J_n \gamma + L\prod I_n K) + K\gamma, \quad (6)$$

where $V_{LP}, V_{FR}, V_{CS}, V_{AP}, V_{IP}, V_K$ are the memory of the fuzzyfication unit, fuzzy rules assertion block, control signal forming block, associative and informative parts, N-dimensional decision field respectively; ; $I_n$, (n=1,…N)- power of universal sets , which the input variables are defined on, J – power of fuzzy sets, L – the number of rules.

Average processing time of the input data stream for the control task in the associative processor with rigid structure can be determined by the formula

$$T_r = T_{LC} + T_{PAMU} + T_{CS} = t_c + t_n^c + 4t_B + t^{min}_n + 2t^{max}_n = \tau(6 + 3\gamma) \quad (7)$$

where $T_{PAMU}$ is PAMU time.

It can be concluded from (7), that the time depends only upon the width of processing words.

The memory cost is evaluated as follows:

$$V_{Ж} = \sum_{n=1}^{N}(V_{LC} + V_{PAMU}) + V_{CS} = \sum_{n=1}^{N}(V_{AP} + V_{IP}' + V_{PAMU}) + V_{CS} = \sum_{n=1}^{N}(\gamma I_n + I_n + 2I_n J_n \gamma + 2NJ_n \gamma) + K\gamma, \quad (8)$$

As it can be seen from the calculations, the processor with flexible structure requires more time and memory.

Therefore, in cases where there is the possibility of changing the control algorithm in the system, it should used flexible processor, in those cases where such a possibility does not exist - the processor with a rigid structure.

### 4. Structural scheme of PAMU

The structure of the associative part of the PAMU, a simplified block diagram of which is shown in Fig. 2, includes:

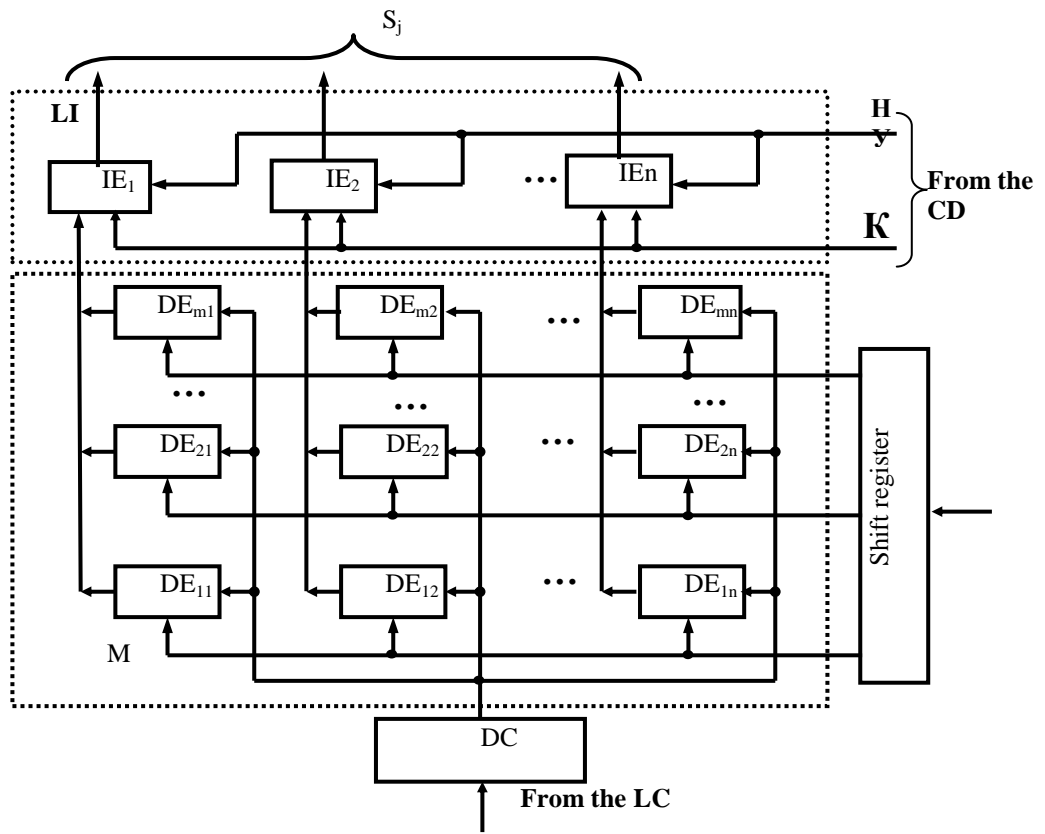

**Fig. 2 Structural scheme of PAMU**

Abbreviations: LC – the linguistic Converter

LI- the line of indication

DE – the deciding element

IE – the indication element

CD – the control device

DC –decoder

1. Matrix M sized by nxm, that interprets the set of states. Where n is a number of etalon sequences, m – maximal length of one of them.
2. m digit shift register that serves to store current state of the automat and controlled by the matrix X.
3. Input decoder performs the transformation of input sequences in unary code, i.e. each input symbol or graduation of criterion corresponds to one of the decoder outputs. In the structural scheme in fig.2 every etalon set corresponds to one row of matrix.

Each element of etalon set stored in the PAMU matrix corresponds to one element of the matrix. The value of the character is determined by "flashing" the matrix from the decoder side, and its serial number in the description is determined the row number of the matrix.

4. The line of the indication (LI) serves for calculation of the similarity functions conjunction.

The developed scheme of PAMU is quite simple and compact. However, its application is effective in the case when descriptions of the equivalent classes of the control situations are received. As in [7], where the conception of the sinterm for the description of the programming language grammars syntactically equivalent terms is introduced.

In addition, when processing different sized serials of the situations they are required to put to the maximal length. Otherwise, the decision procedure should be supplemented with correction algorithms, in the PAMU scheme (Fig. 1) should be introduced a group of valves between the matrix and the line of indication, and in the matrix - the additional valves that serve to fix the end-description sign [10].

The structural scheme allowing handle the input sequences of different sizes is showed on the fig. 3.

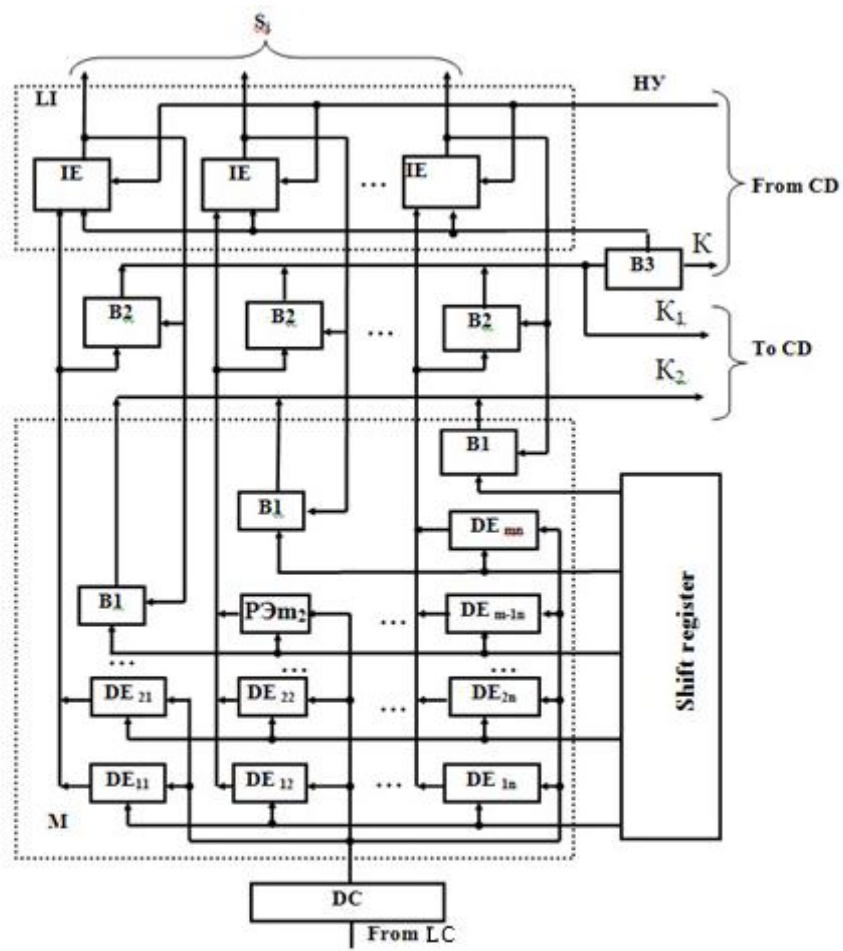

Fig. 3 The structural scheme of PAMU with correction

## 5. Structural scheme of fuzzy associative processor with example of PAMU flashing

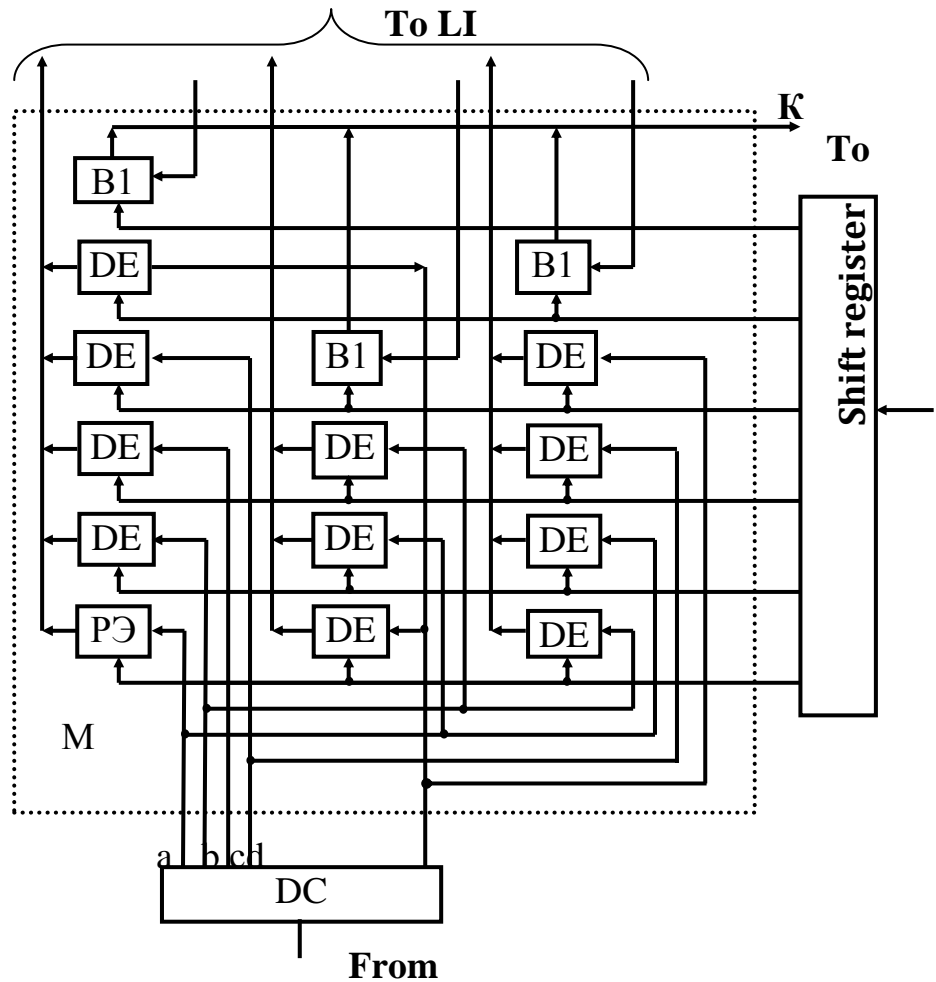

Fig. 4  An example of PAMU flashing

Fig. 4  is an example of "flashing" PAMU matrix for the case of three etalon sets E1 = (a, b, c, d, e), E2 = (e, a, b) and E3 = (b, a, d, e), and Fig. 5 - fuzzy control CPU with a rigid structure built using PAMU and oriented for implementation of the "complete coincidence" algorithm.

PAMU operates as follows. Before the beginning of operation, all indicator elements and the first digit of the distributor
are set to .1.. Upon the arrival of a control signal at the automaton input, the corresponding output bus of the decoder is activated. At the outputs of the matrix elements, for which signals from both the decoder and distributor arrive simultaneously, signals are generated which appear at both the gates B2 and the primary inputs of the coincidence detectors. The secondary inputs of gates B2 are connected to the outputs of the coincidence detectors.

At the first stage of the operation of the unit, all elements of the coincidence detectors are set to "1" and all gates B1 at the matrix output are opened. If there is at least one signal from the PAMU matrix elements, then a signal K1 is generated. In the presence of the signal K1, new values of the coincidence detectors are set according to signal K. If there is a signal at the input of the coincidence detectors, then the detector remains in state "1", otherwise the detector is reset to "0". If there is a signal K1, then simultaneously with the setting of the coincidence detectors,
    the distributor is shifted to the next state.

In the next step of the operation, the next code of the sequence being sequence is presented at the input to the unit and signals are generated at outputs of the PAMU matrix. Signals only appear at the output of the gates for those PAMU columns, for which the coincidence detectors remained in state "1".

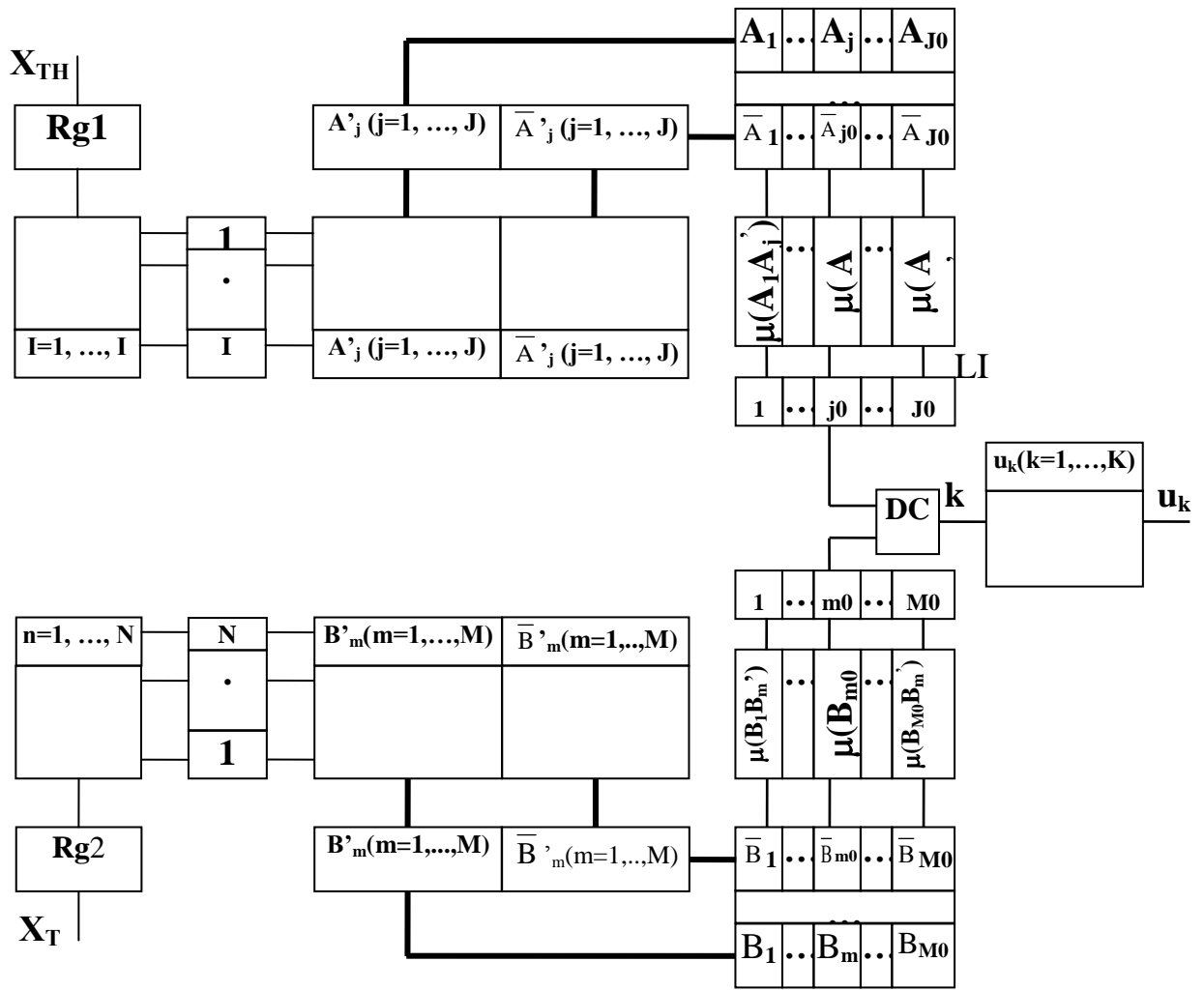

**Fig. 5 The structure of fuzzy associative processor with rigid structure.**

In the presence of the signal K1, new values of the coincidence detectors are set. This process repeats until a signal K2, being an indicator of completion of the comparison, is generated or there is no signal K1 in the next step of the comparison between the input and reference sets. The absence of signal K1 corresponds to non-coincidence between the input and reference sequences, i.e. to the presence of interference in the input sequence. In this case, setting of new values for the coincidence detectors is prohibited which prevents their all being set to 0.

## 6. Conclusion

This paper has focused on the problem of applying associative processor for decision making. Optimization of this design has been done taking into account theoretical and practical issues. Considering theoretical issues, we have decided to focus on associativity as efficient way to implement fuzzy processor. This processor is well-suited for tasks where parameters in the control algorithms don't change frequently. Self-corrected techniques have been applied to tune this processor. Although the structure is quite simple and compact, its application is effective in the case when descriptions of the equivalent classes of the control situations are received. The example PAMU flashing for the case of three etalon sets shows how the "complete coincidence" algorithm works.